# Title: Evidence of Non-Thermal Particles in Coronal Loops Heated Impulsively by Nanoflares

**Authors:** P. Testa[1]*, B. De Pontieu[2,3], J. Allred[4], M. Carlsson[3], F. Reale[5], A. Daw[4], V. Hansteen[3], J. Martinez-Sykora[6], W. Liu[2,7], E.E. DeLuca[1], L. Golub[1], S. McKillop[1], K. Reeves[1], S. Saar[1], H. Tian[1], J. Lemen[2], A. Title[2], P. Boerner[2], N. Hurlburt[2], T.D. Tarbell[2], J.P. Wuelser[2], L. Kleint[2,6], C. Kankelborg[8], S. Jaeggli[8]

**Affiliations:**

[1] Harvard-Smithsonian Center for Astrophysics, 60 Garden st., Cambridge, MA 02138, USA.

[2] Lockheed Martin Solar and Astrophysics Laboratory, 3251 Hanover st., Org. A021S, Bldg.252, Palo Alto, CA, 94304, USA.

[3] Institute of Theoretical Astrophysics, University of Oslo, P.O. Box 1029, Blindern, N-0315, Oslo, Norway.

[4] NASA Goddard Space Flight Center, Greenbelt, MD 20771, USA.

[5] University of Palermo and INAF, Piazza del Parlamento 1, 90134 Palermo, Italy.

[6] Bay Area Environmental Research Institute 596 1st St West, Sonoma, CA, 95476 USA.

[7] W. W. Hansen Experimental Physics Laboratory, Stanford University, Stanford, CA, 94305, USA.

[8] Department of Physics, Montana State University, Bozeman, P.O. Box 173840, Bozeman MT 59717, USA.

*Correspondence to:  ptesta@cfa.harvard.edu

**Abstract**: The physical processes resulting in energy exchange between the Sun's hot corona and its cool lower atmosphere are still poorly understood. The chromosphere and transition region (TR) form an interface region between the surface and the corona that is highly sensitive to the coronal heating mechanism. High resolution observations with the Interface Region Imaging Spectrograph (*IRIS*) reveal rapid variability (~20-60s) of intensity and velocity on small spatial scales (≤500km) at the footpoints of hot and dynamic coronal loops. Comparison with numerical simulations reveal that the observations are consistent with heating by beams of non-thermal electrons and that these beams are generated even in small impulsive (≤30s) heating events called "coronal nanoflares". The accelerated electrons deposit a significant fraction of their energy (≤$10^{25}$erg) in the chromosphere and TR. Our analysis provides tight constraints on the properties of such electron beams and new diagnostics for their presence in the non-flaring corona.

**Main Text:**

Though it is established that the magnetic field plays a major role in the energetics of the bright corona, determining the details of the physical mechanisms that heat the solar corona remains one of the outstanding open issues in astrophysics.  There are several candidate physical processes for heating the corona, including dissipation of magnetic stresses via reconnection, and dissipation of magnetohydrodynamic waves *(1,2,3)*. In many heating models, the energy release is characterized by small spatial and temporal scales. For instance, in the "nanoflare" model, random photospheric motions lead to braiding or shearing of magnetic field lines and to reconnection which yields impulsive heating of the coronal plasma *(4,5)*. Several statistical studies of large numbers of solar flares (*6-8*) have suggested that the mechanisms producing flares are likely similar within a large range from micro- to X-class flares. If nanoflares behave as a scaled down version of larger flares, particles accelerated in the corona by reconnection processes could play a significant role in the heating of plasma even in absence of large flares. Hard X-ray observations of microflares ($E\sim 10^{27}$erg) in active regions reveal the presence of non-thermal particles *(8,9)*, but nanoflare size events ($E\sim 10^{24}$erg) are not currently accessible to hard X-ray studies due to limited sensitivity. As a result, the properties and generation of non-thermal particles in the solar atmosphere and their role in quiescent coronal heating remain poorly known.

The observational tracers of the coronal heating are elusive because the corona is highly conductive, washing out the signatures of heating release. However, the emission of the TR, where the temperature steeply increases to MK values in a narrow layer ($\sim$1-3 $\times 10^8$ cm), is instead highly responsive to heating since its density, temperature gradients and spatial dimensions change rapidly during heating events *(10-12)*. This is the also the case for coronal heating events where non-thermal electrons are produced that lose most of their energy through collisions with dense chromospheric and transition region plasma (*thick-target* model, *13-14*).

Recently the High-resolution Coronal Imager (*Hi-C*) rocket experiment *(15)* provided high cadence ($\sim$5s) extreme ultraviolet (EUV) imaging observations of the corona at the highest spatial ($\sim$200 km) resolution to date. *Hi-C* observations of the upper TR at the footpoints of hot (>4 MK) coronal loops (the "moss") have revealed rapidly variable emission, consistent with coronal nanoflares due to slipping reconnection *(12)*. However, the lack of spectral information in *Hi-C* data precludes an accurate determination of the plasma properties (e.g., plasma flows at different temperatures) and therefore prevents a detailed study of the physical processes at work.

The Interface Region Imaging Spectrograph (*IRIS*, *16*), launched in June 2013, is exceptionally well suited to investigate the response of the lower atmosphere to heating events, as it provides imaging and spectral observations of chromospheric and TR emission at high spatial, temporal and spectral resolution. We study short lived brightenings at the footpoints of hot dynamic loops (Fig. 1, 2, Movies S1-S2), and analyze the *IRIS* observations, together with coronal observations with the Atmospheric Imaging Assembly (*AIA, 17*) onboard the Solar Dynamics Observatory (*SDO*, *18*). We couple the data analysis to advanced numerical modeling of impulsively heated loops, to diagnose the properties of the heating, the mechanisms of energy transport, and the dynamic plasma processes. We show that the interaction of non-thermal electrons, accelerated in small ($E \lesssim 10^{25}$erg) heating events, with the lower atmosphere can reproduce the IRIS observations.

We observe footpoint brightenings in *IRIS* chromospheric and TR lines (e.g., MgII k 2796.4 Å, CII 1335.78 Å, Si IV 1402.77 Å, formed around log($T_{max}$[K])~4.0, 4.5, and 4.9 respectively; *19*), as well as in upper TR lines with *AIA* (e.g., 171 Å, 193 Å passbands, which are mostly sensitive to 0.7-1.5MK emission; *19*). The moss brightenings are clearly associated with heating of the overlying loops, which become brighter minutes later (Movie S2). The brightenings show a typical duration of 20-60 s (Fig.2), and their intensity variations span more than two orders of magnitude, especially for Si IV (Fig. 3, Fig. S2, S3). For a few brightenings occurring at the location of the *IRIS* spectrograph slit we also obtained far-ultraviolet (FUV) and near-ultraviolet (NUV) spectra. The Si IV spectra (Fig. 3) indicate that many of the brightenings are associated with modest upflows (blue-shifts, i.e., v<0) with typical velocities of ~15 km/s.

State-of-the-art physical models are necessary to interpret the observed spectral and temporal evolution of the chromospheric and TR plasma. We use the RADYN hydrodynamic model (*20,21*) of plasma confined in a loop, which includes non-Local Thermodynamic Equilibrium (non-LTE) radiative transfer, and also allows us to model heating by non-thermal electron beams (*19*). We simulate the chromospheric and TR response to impulsive (10-30 s duration) heating with total energy ~$10^{24}$-$10^{25}$ erg. As estimated from the observations (*19*) we assume the loop to be of half-length and cross-section of $10^{10}$ cm and $5 \times 10^{14}$ cm$^2$ respectively, and initially of low coronal density (log ($n_e$ [cm$^{-3}$])~7.5), and temperature (log ($T_{max}$ [K])~6). We consider two different heating mechanisms: 1. a power-law distribution of high-energy non-thermal particles, 2. localized coronal heating without accelerated particles, resulting in thermal conduction as the primary mechanism of energy transport from the corona to the lower atmospheric layers. For the simulations with non-thermal electron beams we explored the effects of the energy flux, and of the low energy cutoff value $E_c$ (*19*).

The synthetic *IRIS* and *AIA* observables from the model show that the duration of the chromospheric and TR brightenings is generally very similar to the duration of the heating event (Fig. 4, *19*). Therefore, the temporal variability of the TR/chromospheric emission is an excellent direct diagnostic of the temporal properties of the heating for short events.

The observable that discriminates most efficiently between the beam heating models and the conduction models is the Doppler shift in the Si IV emission. Models with beam heating generally predict blue-shifts in Si IV, with typical associated velocities analogous to the observed ones. The blue-shifts are due to the fact that the energetic particles initially deposit a large fraction of their energy in the chromosphere, locally heating it, with the subsequent pressure pulse triggering both upflows for the hotter plasma above this layer, and downflows for the cooler plasma below. The depth of the layer where most of the beam energy is deposited sensitively depends on $E_c$ (*19,22*). In the absence of accelerated particles, thermal conduction transports the energy, originally deposited in the corona, to the lower atmosphere, and the TR/chromosphere is pushed down, leading only to redshifted emission in the initial impulsive phase.

Therefore, the blue-shifts observed in Si IV for many footpoint brightenings are incompatible with heating transported by thermal conduction or by of Alfven wave

*(19,23)*, but are compatible with beams of non-thermal electrons depositing energy directly in the chromosphere and lower transition region.

The models with electron beams also naturally reproduce a larger range of brightening intensities observed in Mg II, Si IV, *AIA* 171 Å and 193 Å. In particular, the Si IV emission is formed at heights where these nanoflare-sized beams naturally deposit most of their energy: large brightenings are observed when the combination of electron energy cutoff and total energy flux result in high densities at temperatures that contribute to the Si IV lines. In the conductive case the brightenings in layers ranging from the chromosphere to the upper TR, are more tightly coupled as they are produced as a direct consequence of the energy conducted down from the corona (*19*). In the beam heating models, the heating of the lower chromospheric layers is partially decoupled from the heating of the higher TR layers because non-thermal particles of different energies will deposit energy in different layers. In the cases of a hard distribution ($E_c$ ~15-20 keV), for the considered total energies (which are significantly lower than in larger flares), the energy is deposited in the low chromosphere and does not produce observable transient brightenings in the TR emission (*19*). Softer distributions with $E_c$ < 10 keV generally deposit much of their energy in the corona, thus leading to results analogous to heating by conduction (*19*).

The combination of IRIS observations and modeling can thus provide a novel diagnostic of non-thermal particle heating in small coronal heating events or nanoflares. The high sensitivity of the TR emission to the beam parameters implies that *IRIS* observations help constrain the cutoff value $E_C$ of the non-thermal electron distribution. Determining $E_C$ from observations is of crucial importance to derive the non-thermal energy content, because most of the beam energy is found around $E_C$, and to constrain mechanisms of particle acceleration *(22, 24-25)*. Hard X-ray observations are the main diagnostic tool to study non-thermal electron distributions *(26)*, because the bremsstrahlung process is well understood *(27)*. However, they are intrinsically limited in their ability to constrain $E_C$, mainly because the thermal emission typically dominates the emission at lower energies (≲20 keV; *13*). Our *IRIS* spectral observations of footpoint brightenings show that the heating occurs on small timescales (≲ 30s), that each event is characterized by total energy ≲ $10^{25}$ erg, and that $E_C$ ~10 keV. We note that a few brightenings also show redshifts or no shifts and they might be explained by either conduction or by beams with low $E_C$ value (*19*).

*IRIS* chromospheric and TR observations offer a unique opportunity to investigate non-thermal electron beams, as an alternative to hard X-ray observations. The *IRIS* beam diagnostics have higher sensitivity than hard-X-ray observations, allowing us to study nanoflare-size heating events at high spatial resolution, and to constrain the low energy tail of the beam distribution. Systematic studies of *IRIS* and *AIA* observations alongside numerical models will help determine the prevalence of beam heating in active regions, as well as its role in the energetics of non-flaring plasma. Our results are also relevant for space physics and astrophysics, where particle acceleration is thought to play a major role in many phenomena (e.g., *28-30*).

1. **References and Notes:**
1. J. Klimchuk, *Solar Phys.* **234**, 41 (2006)
2. F. Reale, *Living Rev. Solar Phys.* **7**, 5 (2010)
3. A. van Ballegooijen et al., *Astrophys. J.* **736**, 3 (2011)
4. E.N. Parker, E.N., *Astrophys. J.* **330**, 474 (1988)
5. E.R. Priest, J.F. Heyvaerts, A.M. Title, *Astrophys. J.* **576**, 533 (2002)
6. N. Crosby, M. Aschwanden, B. Dennis, *Solar Phys.* **143**, 275 (1993)
7. U. Feldman et al., *Astrophys. J.* **460**, 1034 (1996)
8. S. Christe et al., *Astrophys. J.* **677**, 1385 (2008)
9. I. Hannah et al., *Space Sci. Rev.* **159**, 263 (2011)
10. L. Fletcher, B. De Pontieu, *Astrophys. J. Letters* **520**, 135 (1999)
11. P.C.H. Martens, C.C. Kankelborg, T.E. Berger, *Astrophys. J.* **537**, 471 (2000)
12. P. Testa et al., *Astrophys. J. Letters* **770**, 1 (2013)
13. J. Brown, et al., *Astronomy & Astrophys.* **508**, 993 (2009)
14. M. Varady, et al., *Astronomy & Astrophys.* **563**, 51 (2014)
15. J. Cirtain, et al., *Nature* **493**, 501 (2013)
16. B. De Pontieu, et al., *Solar Physics* **289**, 7, 2733 (2014)
17. J. Lemen et al., *Solar Physics* **275**, 17 (2012)
18. D. Pesnell et al., *Solar Physics* **275**, 3 (2012)
19. Supplementary Material.
20. M. Carlsson, R. Stein, *Astrophys. J.* **481**, 500 (1997)
21. J. Allred et al., *Astrophys. J.* **644**, 484 (2006)
22. G. Holman et al., *Space Sci. Rev.* **159**, 107 (2011)
23. L. Fletcher, & H. Hudson, *Astrophys. J.* 675, 164 (2008)
24. V. Zharkova et al. *Space Sci. Rev.* **159**, 357 (2011)
25. V. Petrosian, V., & S. Liu, *Astrophys. J.* **610**, 550 (2004)
26. E. Kontar et al., *Space Sci. Rev.* **159**, 301 (2011)
27. H. Koch, & J. Motz, *Review of Modern Physics* **31**, 920 (1959)
28. R. Blandford, & J. Ostriker, *Astrophys. J. Letters* **221**, 29 (1978)
29. A. Bell, *Month. Not. Royal Astron. Soc.* **353**, 550 (2004)
30. V. Petrosian, A. Bykov, *Space Sci. Rev.* **134**, 207 (2008)

**Acknowledgments:** IRIS is a NASA Small Explorer mission developed and operated by LMSAL with mission operations executed at NASA Ames Research Center and major contributions to downlink communications funded by the Norwegian Space Center (NSC, Norway) through an ESA PRODEX contract. This work is supported by NASA under contract NNG09FA40C (IRIS) and the Lockheed Martin Independent Research Program, the European Research Council grant agreement No.291058, and contract 8100002705 from Lockheed-Martin to SAO.

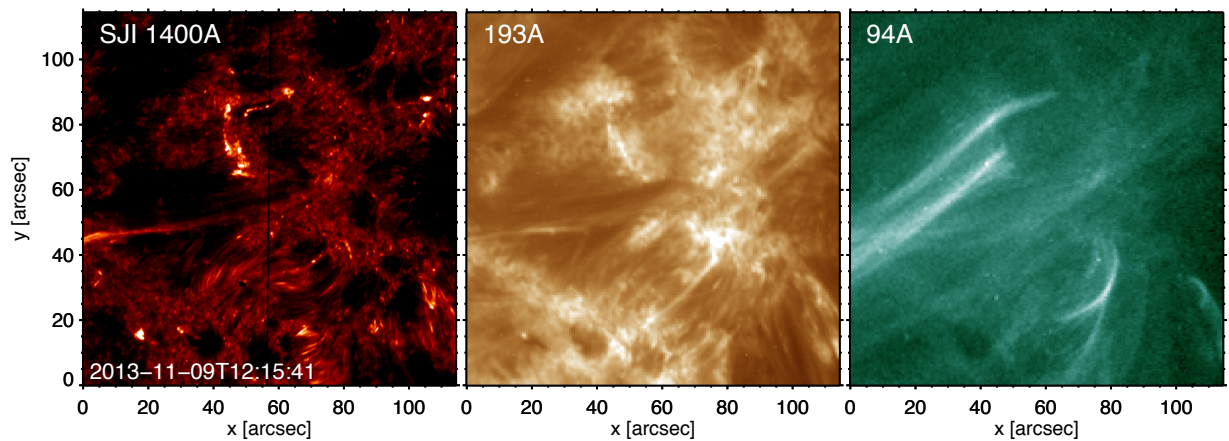

**Fig. 1.** *Imaging observations of moss brightenings associated with coronal loop heating.* Coronal and TR images of active region 11890, on 2013-11-09. The *IRIS* slit-jaw image (SJI) in the 1400 Å passband, and the *AIA/SDO* 193 Å image are dominated by TR emission, while the 94 Å images, is dominated by hot coronal emission (*20*). The brightenings we focus on occur around x=40-65 and y=60-90 (see also Movies S1 and S2; in *20* we also discuss brightenings that occur around x~55 and y=20-25).

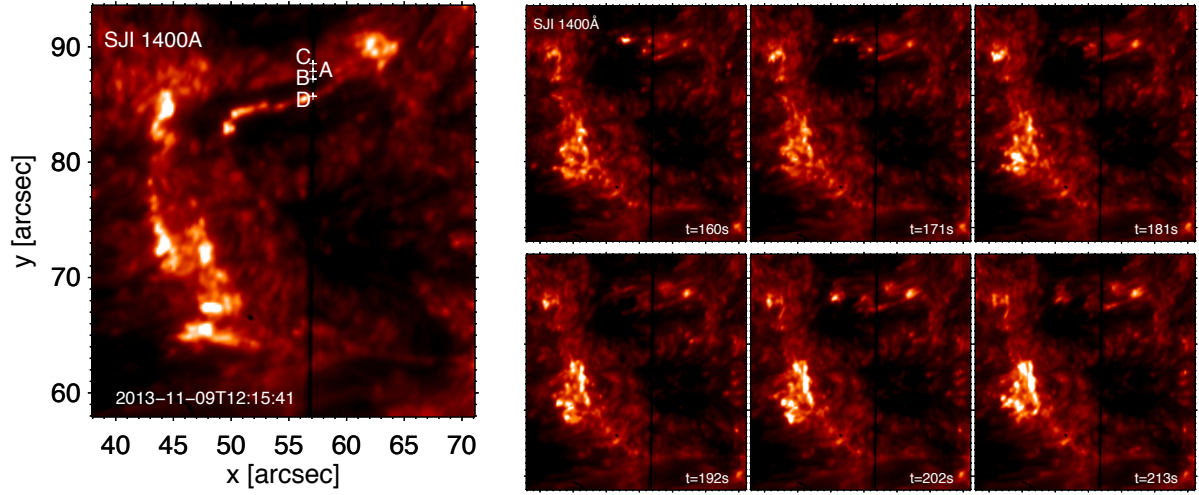

**Fig. 2.** *Temporal evolution of TR emission for observed brightenings is dominated by timescales of order 10-30s.* Left panel: *IRIS* 1400 Å SJI of a region where most of the footpoint chromospheric and TR brightenings occur. A few locations are marked for which the temporal evolution of the chromospheric/TR/coronal emission is shown in Figure 3. Right panel: *IRIS* SJI observations of the same subregion shown in the left panel, at different times (see also *20* and Fig. S1-S2). Times are in seconds from the start of the IRIS observations, 2013-11-09 12:04:16UT.

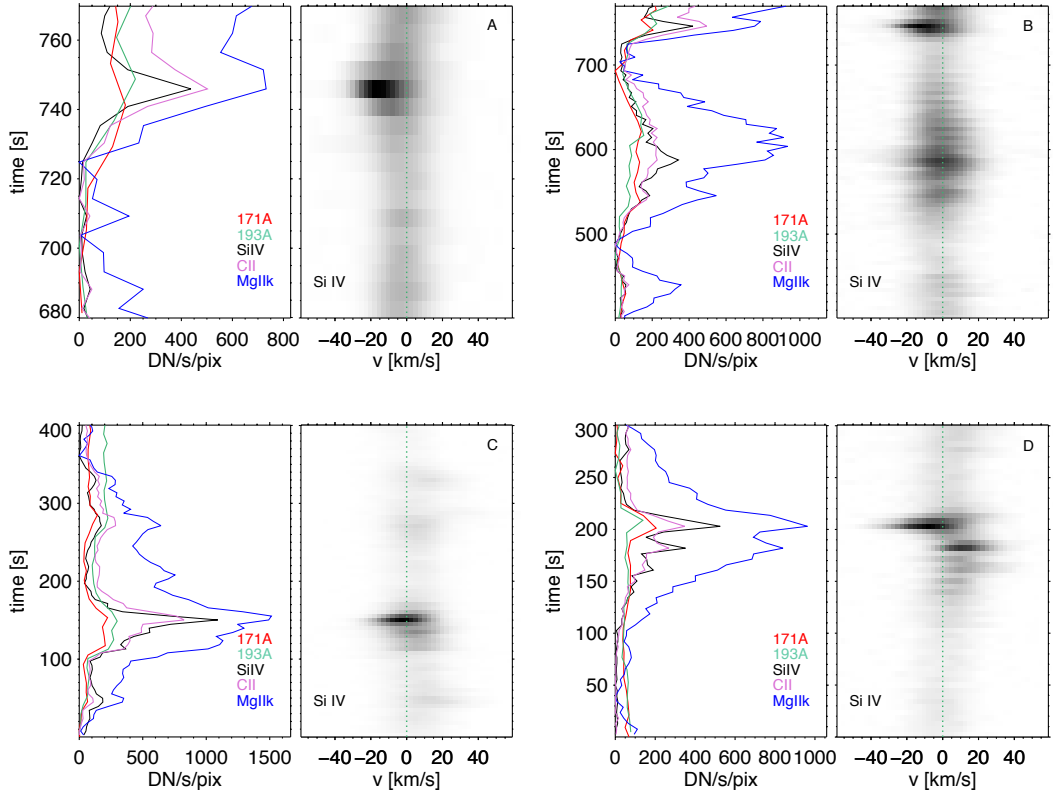

**Fig. 3.** *Temporal evolution for observed brightenings of line/passband intensities and Si IV spectra shows a wide range of upper and lower transition region response.* For the locations marked (A-D) in Figure 2 we show: (left panel) the intensities vs. time (with the minimum value of the time series subtracted) of FUV/NUV emission lines (Si IV 1402.77 Å, C II 1335.71 Å, Mg II k 2796.4 Å) observed with *IRIS*, and in *AIA* EUV narrow bands (171 Å, 193 Å); the Si IV spectra vs. time (right panel), where the wavelengths have been converted to Doppler velocities (the dotted line marks v=0). Negative velocity values (i.e., blueshifts) correspond to upflows, positive velocities (i.e., redshifts) to downflows.

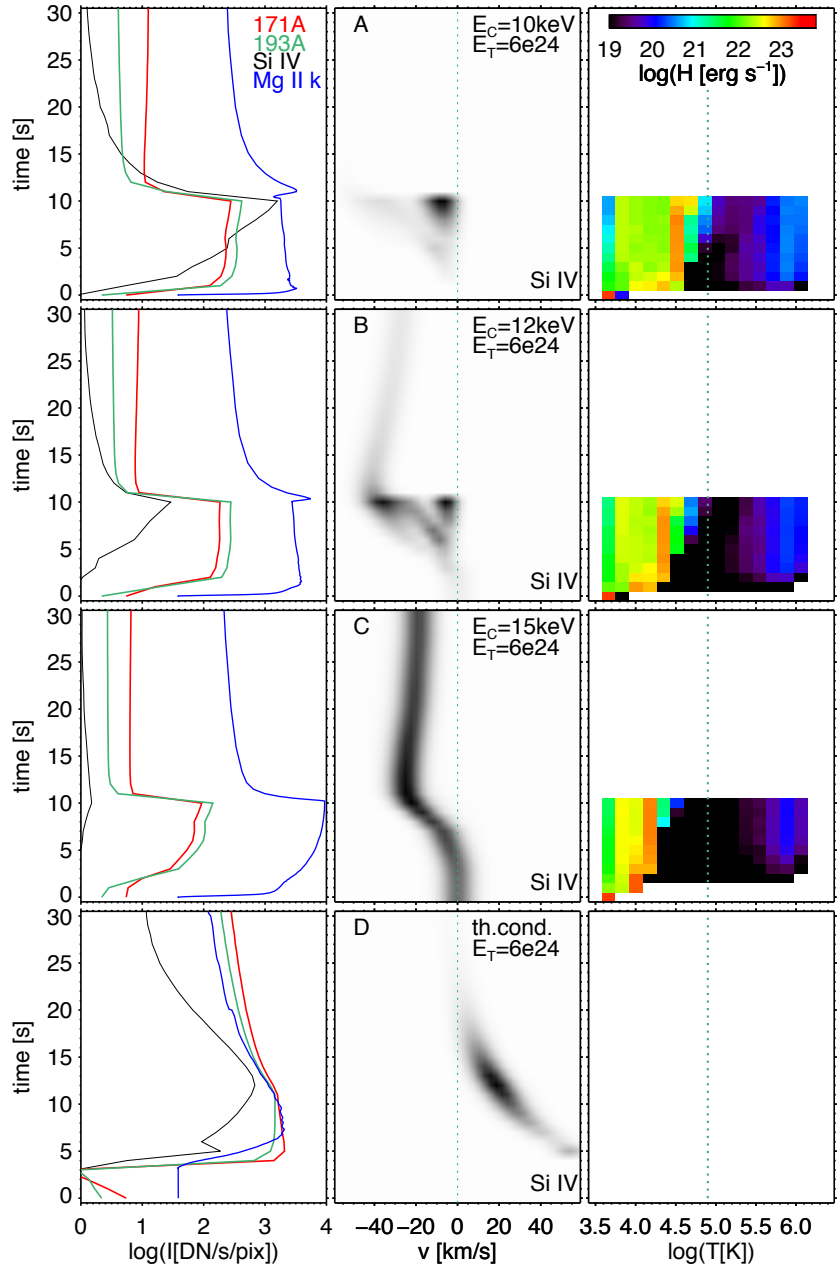

**Fig. 4**. **Loop *models with impulsive heating from non-thermal electron beams reproduce properties of observed loop footpoint brightenings.*** We show synthetic *IRIS* and *AIA* emission from RADYN simulations with heating duration of 10s, total energy $E_T = 6\times10^{24}$ erg, and: energy cutoff $E_C$, equal to 10 keV (A), 12 keV (B), 15 keV (C); panel D shows a model without accelerated particles. For each model we show: *IRIS* and *AIA* emission lightcurves (left), Si IV spectra vs. time (middle), and, for beam models, the heating rate as a function of temperature and time (right; the dotted line marks the temperature of maximum formation of the Si IV line), due to the interaction of the energetic particles with the dense plasma in the lower atmosphere.

**Supplementary Materials:**

Materials and Methods

Figures S1-S6

Movies S1-S10

References (*31-74*)

**Supplementary Materials:**

**S1. Materials and Methods:**

**S1.1 Observations and Data Reduction**

We analyze *IRIS* and *SDO/AIA* observations of active region (AR) 11890 on November 9 2013, from 12:04 to 12:18UT (Fig.1, and Movies S1, and S2).

The *IRIS* dataset we use is a high cadence sit-and-stare observation with large field of view (0.33"x120" for spectral data, 120"x120" for SJI), and exposure time of 4 s. The FUV and NUV spectra (flare line list) are taken at a cadence of ~5.2 s, and the SJI in 1400 Å and 2796 Å passbands are taken at ~10.5s cadence. In this dataset both the FUV and NUV spectra have $\Delta\lambda$~0.025 Å. We use calibrated level 2 data, which have been processed for dark current, flat field, and geometrical corrections. The wavelength scale has been corrected for small thermal drifts by using the routine iris_orbitvar_corr_l2.pro available in SolarSoft. We use the neutral line of O I at 1355.6 Å, which is expected to have intrinsic velocity of less than 1 km/s, to correct the absolute wavelength scale in the FUV.

For *AIA*, we use level 1.5 data obtained by applying the *aia_prep* routine available on SolarSoft to process level 1 data (which are already processed to apply bad-pixel removal, despiking, and flat-fielding), performing image registration (coalignment, and adjustments of roll angle, and plate scales between the image in different passbands). The data we use are from standard observing series with cadence of 12s, and are characterized by exposure times of 2 s in the 171 Å and 193 Å channels and of 2.9 s in the 94 Å passband.

We coalign the *IRIS* and *AIA* data to remove residual small shifts and small relative roll angle between *IRIS* and *AIA* images. The *AIA* images (~0.6"/pixel) have been rescaled to the *IRIS* SJI spatial resolution (~0.166"/pixel), and corrected for solar rotation. We create a time series in each *AIA* band, by taking the corresponding image closest in time to the *IRIS* observing times. We then use cross-correlation to coalign *IRIS* 1400 Å SJI images with *AIA* 1600 Å images, which typically are, among *AIA* images, the ones with morphology most similar to *IRIS* 1400 Å SJI.

The *AIA* data allow us to estimate the half-length of the loops overlying the footpoints which undergo rapid variability, to be of the order of $10^{10}$ cm. Many of the *IRIS* brightenings are found to be of spatial extent close to the *IRIS* resolution, and similar to what we had found for the *Hi-C* moss brightenings *(12)*, therefore, for the loop modeling described in S1.2, we assumed a loop cross-section of ~5 x $10^{14}$ $cm^2$.

In our analysis we focus on *AIA* 171 Å, 193 Å, 94 Å data, and on the *IRIS* Si IV 1402.77 Å, and MgII k 2796.4 Å spectral lines and SJI 1400 Å images. In typical coronal conditions, the 171 Å and 193 Å *AIA* passbands are dominated by the emission of the Fe IX ($\log(T_{max}[K])$~5.95) and Fe XII ($\log(T_{max}[K])$~6.2) respectively, while the 94 Å passband includes a strong Fe XVIII line ($\log(T_{max}[K])$~6.85) but is also sensitive to ~1-2MK plasma *(31-34)*. The Fe XVIII emission is however dominant when hot plasma is present, like in the core of hot active regions *(35,36)*. The Si IV lines in the IRIS spectra have a temperature of peak formation $\log(T_{max}[K])$~4.9. The Mg II lines are formed around log(T[K])~4.0 with various spectral features formed over a wide height range from the upper photosphere (k1/h1), middle chromosphere (k2/h2) to the upper chromosphere (k3/h3) (*37*).

In Supplementary Movie S1 we show the *IRIS* (SJI 1400 Å, and 2796 Å) and *AIA* (94 Å, 193 Å) time series of imaging observations of TR and coronal emission in AR 11890 (2013-11-09 12:04-12:17UT). The footpoint brightenings are visible both in *IRIS* SJI and *AIA* 193 Å. The 94 Å images show that these brightenings are at the footpoints of hot variable loops (≳4MK). Suppl. Movie S2 shows the evolution of AR 11890, from 11:45UT to 12:45UT, as observed by *AIA* in the 94 Å, 193 Å, 171 Å, and 304 Å passbands. The 304 Å emission is dominated by He II 303.8 Å lines (log($T_{max}$[K])~4.9, *32*). This movie shows that the footpoint brightenings are observed in the *AIA* 193 Å, *AIA* 171 Å, and *AIA* 304 Å passbands.

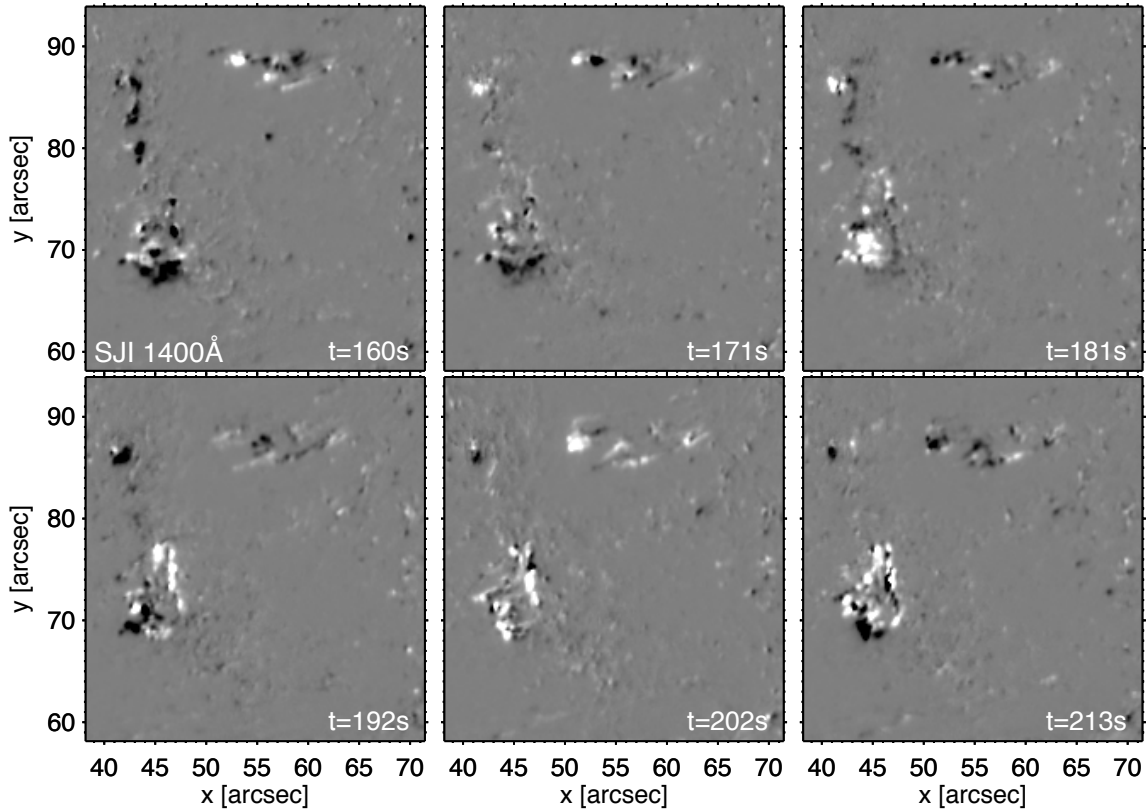

**Fig. S1**. *Spatial and temporal variations of TR emission for observed brightenings.* We show running difference (image at time $t_{i-1}$ subtracted from the image at time $t_i$) for *IRIS* 1400 Å SJI in the same region shown in Fig. 2, and for the same times.

Figure S1 and S2 show more details of the temporal variability during the footpoint brightenings. We calculate the running difference (i.e., for a time $t_i$ we subtract the image at time $t_{i-1}$ from the image at time $t_i$) for the *IRIS* 1400 Å SJI, and in Fig.S1 we show it for the same subregion and same times as the images of Fig. 2. Locations with intensity increase are in white, and locations where the intensity decreases are shown in black. The running difference maps show that the variability occurs on small spatial and temporal scales.

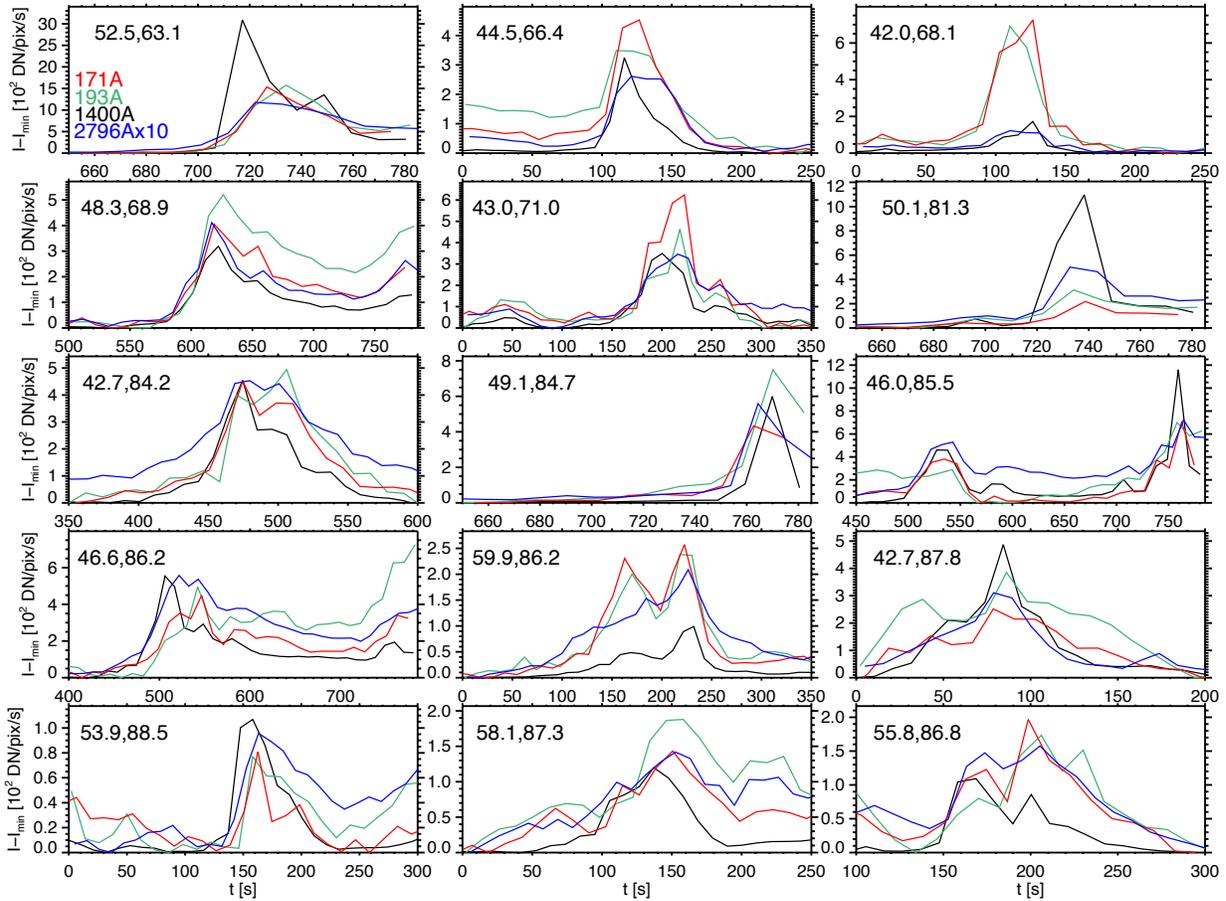

**Fig. S2**. *Temporal evolution of chromospheric and TR emission for observed brightenings from SJI and AIA imaging data*. We show lightcurves in the *IRIS* SJI 2796 Å (multiplied by 10) and 1400 Å passbands, and in the AIA 193 Å and 171 Å passbands, for a sample of brightenings. In each plot we report the coordinates of the location of the brightening (x,y, in arcsec, using the same reference of Figures 1,2, and S1).

In Figure S2 we show lightcurves in the *IRIS* 1400 Å and 2796 Å SJI, as well as for *AIA* 171 Å and 193 Å passbands, in a sample of footpoint locations undergoing short-lived brightenings. In the plots we multiply the 2796 Å SJI intensities by 10. We note that the effective area for the SJI is smaller than the effective area for the spectrograph (SG). In particular, the effective area of SJI at 1402 Å (i.e., at the wavelength of the Si IV line) is about 3 times smaller than the SG effective area at the same wavelength, while the SJI effective area at 2796 Å is about 40 times smaller than the corresponding SG effective area (*16*). Therefore, when comparing the observed lightcurves with the predictions from simulations (Fig.4 and sections S1.2 and S2), these scaling factors need to be taken into account as well as the fact that SJI contains other contributions other than the Si IV and Mg II k line respectively. In particular, while for the *IRIS* 1400 Å SJI the Si IV emission is expected to dominate the emission, during these brightenings, for the *IRIS* 2796 Å SJI contributions from lines other than the Mg II k line, and from continuum are very significant.

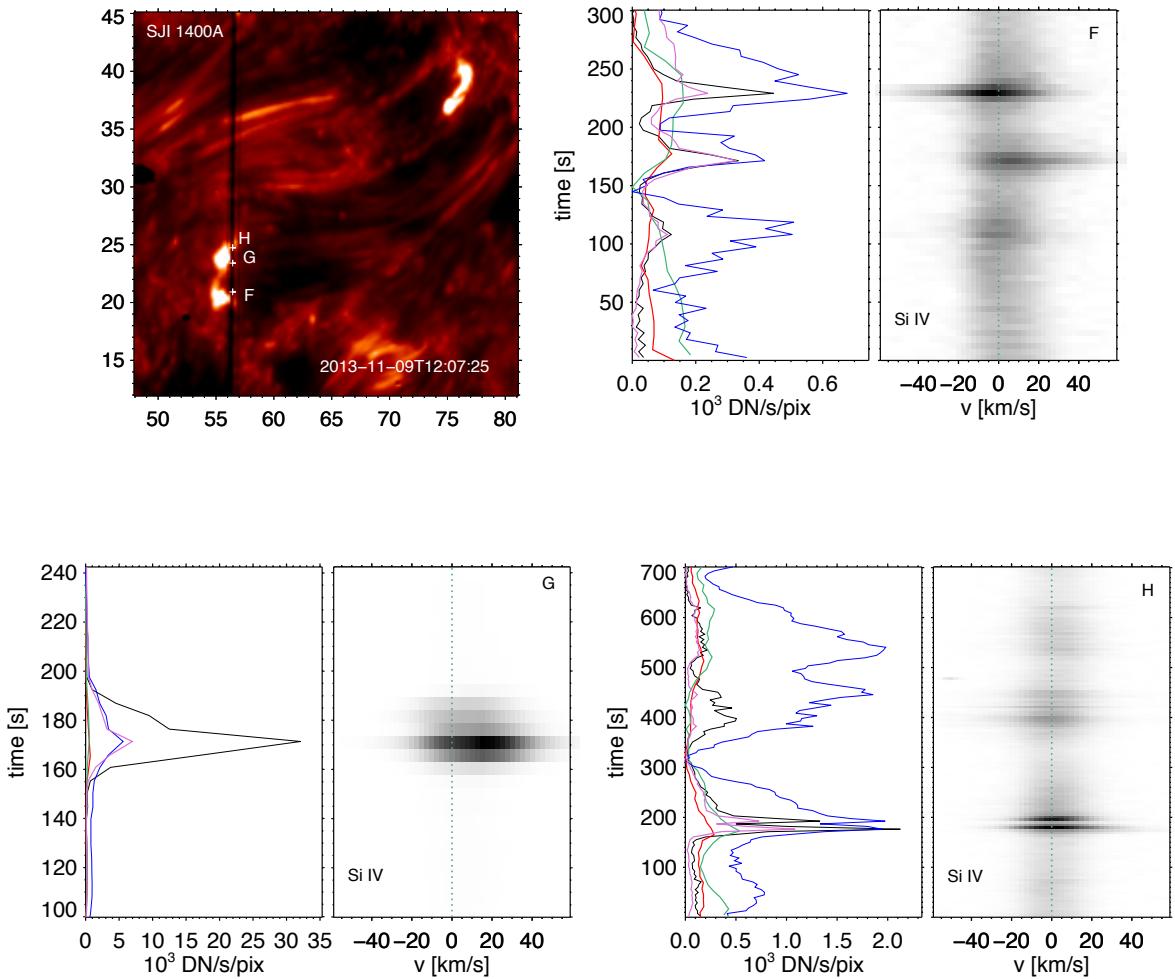

**Fig. S3**. *Temporal evolution of chromospheric and TR emission for brightenings at footpoints of hot loops observed in the southern portion of the IRIS field of view*. The top left panel shows an *IRIS* SJI 1400 Å image at a time when the brightening occurs. The other three panels show lightcurves of TR and chromospheric emission, and Si IV spectra vs. time, for these brightenings, in the same format as Fig. 3 of the main text.

Movies S1 and S2 shows that strong footpoint brightenings occur also at the base of another set of shorter hot loops (with loop length ~ 2 x $10^9$ cm), that brighten around 12:07-12:10UT in the *AIA* 94 Å passband (around x~55 and y=20-25 in Fig.1). We show in Fig. S3 the temporal evolution of these brightenings in AIA passbands and IRIS spectral lines, analogously to Fig. 3. Also for these brightenings the lightcurves show a wide range of intensity values, especially in Si IV. The Si IV spectra of the brightenings also show a range of behavior, including blueshift, redshift and lack of appreciable Doppler shift.

### S1.2 Numerical Simulations

We have developed models of how nanoflares heat the chromosphere and transition region to test if they could be the source of the observed loop footpoint variability. In standard flare theory (e.g., *38-41*), impulsive energy release during magnetic reconnection leads to acceleration of particles to high energies and/or plasma heating in the corona. These non-thermal particles (electrons and ions) are confined by the relatively strong field to travel along magnetic field lines. As they move down, they collide with and heat the ambient plasma. The heated atmosphere produces a wide range of emission from EUV to infrared (e.g.,*42*). Most flares are electron-dominated as indicated by their commonly detected bremsstrahlung hard X-rays, while gamma-ray-producing ions are only found in intense major flares (*43*) and their energy budgets are comparable to those of electrons (*44*).

To simulate how the chromosphere and transition region respond to nanoflare heating, we have used the RADYN code (*22-23*, *45-47*). RADYN solves the equations of radiation hydrodynamics in one spatial dimension, which is assumed to lie along the axis of a magnetic flux tube. The equations are formulated on an adaptive grid (*48*) such that large gradients can be resolved with a rather small number of spatial grid cells. In the simulations presented here we use 191 spatial grid cells and the grid distance varies from 0.5 m in the transition region and strong shocks to 2 Mm at the top of the computational domain.

A key strength of RADYN is its ability to capture dynamics in the chromosphere where non-LTE, optically-thick radiative transfer significantly contributes to the energy balance. RADYN solves the equations of hydrodynamics coupled with the non-LTE population densities for numerous atomic levels in hydrogen, helium, and singly ionized calcium and magnesium. These ions where chosen because they produce transitions which are energetically important in the chromosphere. RADYN solves the radiative transfer equation for 46 atomic transitions each of which is resolved with up to 100 frequency points and 5 directional rays.

In addition to the optically-thick transitions modeled in detail as described above, there are numerous optically-thin transitions which are important to the energy balance in the transition region and corona. The effects of these are included using an overall radiative loss function which is generated using the CHIANTI database (*49,50*).

RADYN includes thermal conductivity using the classical Spitzer formula. However, during flare simulations the temperature gradient in the transition region can become so large that Spitzer conductivity predicts fluxes that exceed the electron free-streaming rate. When this occurs the conductive flux is capped at the free-streaming rate as described by (*51*).

Our model flux tubes are rooted in the sub-photosphere and stretch out to include the chromosphere, transition region and corona.  They were constructed by adding a prescribed amount of heating into the corona and the sub-photosphere and allowing RADYN to evolve the loops until a state of near equilibrium was reached. By varying the coronal heating rate and loop length, we have generated several flux tubes with coronal temperatures and densities ranging from hot, dense loops (5 MK, $1 \times 10^{10}$ cm$^{-3}$) to cool, low density loops (1 MK, $3 \times 10^7$ cm$^{-3}$). As described in S1.1, we used the *AIA* and *IRIS* observations to estimate the loop half-length and cross-section ($10^{10}$ cm, $5 \times 10^{14}$ cm$^2$ respectively) to be used for the models.

The boundary conditions are specified as follows. The lower boundary is in the sub-photosphere, and it uses an "open" boundary allowing waves to pass through. At the outer boundary we have implemented a reflecting condition. Waves hitting the top will reflect back mimicking waves entering from the other side. All gradients at the loop top are set to zero, as would be expected at the top of a symmetric loop.

RADYN has been modified, as described by Abbett & Hawley (*52*) and Allred et al. (*53*), to include the effects of heating due to non-thermal electron beams. More recently, Allred et al. (*54*) have enhanced RADYN to simulate the transport of accelerated electrons through the solar atmosphere using the Fokker-Planck kinetic theory. This is a step further from a similar prototype hybrid hydrodynamic-kinetic simulation (*55*), which used the same particle transport module (*56*) plus a stochastic acceleration module (*25*) but without radiative transfer calculation. The Fokker-Planck kinetic theory describes the evolution of the electron distribution function as they travel through the flux tube. Since the electrons are confined to spiral along magnetic field lines, their distribution function is fully described by their position along the flux tube, energy, and pitch-angle.  The electron transport time is much faster than the duration of nanoflare heating, meaning transients will quickly dissipate and the distribution function will reach an equilibrium state, so we assume time-independence in solving the Fokker-Planck equation.

We solve the Fokker-Planck equation using the technique described by McTiernan & Petrosian (*56*). Specifically, we model the precipitation of beams of electrons injected at the loop top. The technique is fully relativistic and includes the effects of energy loss and pitch angle scattering due to Coulomb collisions and synchrotron emission, and pitch angle changes due to magnetic mirroring.  In order to accurately capture these effects, we found it necessary to construct a grid with 120 points in energy space, 30 in pitch angle, and 191 in position. The energy grid is spaced non-linearly such that energies near the electron cutoff are much more finely resolved than those at very high energy. Likewise, the pitch-angle grid is spaced to resolve angles nearly aligned with the spatial axis.

By analyzing many thousands of flare observations, Christe et al. (*8*) found electrons accelerated during microflares have a power law energy spectrum and derived an average spectral index, δ. By extension, we assume that nanoflare-accelerated electrons also have a power law distribution and have chosen a spectral index of 5, similar to what they found. Even for large flares, it is typically only possible to place an upper-limit on the cutoff of the electron energy spectrum. This is because thermal X-ray emission usually dominates the non-thermal emission in the range where the energy cutoff occurs. Christe et al. (*8*) found that 20 keV is a typical upper limit, so in our parameter study, we have varied the cutoff energy from 5 – 20 keV.

For simulations without non-thermal particles, the impulsive heating is released in the corona, and the energy is transported to the lower atmospheric layers by thermal conduction. These RADYN simulations with heating by conduction have also been compared to results of simulations performed with the Palermo-Harvard loop model (*57,58*).

For each RADYN simulation we use the hydrodynamical variables at each time-step as input to synthesize observables. The resonance lines of singly ionized magnesium (the Mg II h & k lines) are optically thick and a full radiative transfer non-LTE calculation is required. We use a modified version of the code RH (*59*) that includes angle-dependent partial frequency redistribution (PRD) using the fast approximations of Leenaarts et al. (*60*). We use the 10 level plus continuum Mg II model atom of Leenaarts et al. (*37*) with a magnesium abundance of 7.6 (*61*) on the standard logarithmic scale where the abundance of hydrogen is 12.

The optically thin emission in the TR and coronal line/passbands are calculated (as in *62*) using emissivity functions from CHIANTI (*42, 43*), and assuming coronal abundances (*63*) and ionization equilibrium. For each line/passband we fold the emission through the instrument responses (*16,31*) to obtain intensities in units of DN s$^{-1}$ pix$^{-1}$ (for *IRIS* we synthesize intensities using the spectrograph effective areas).

### S2. Simulation Results

Using RADYN simulations we have explored the effects of: (a) energy cutoff value $E_C$ characterizing the electron power-law distribution, for beam simulations; (b) total energy, $E_T$; (c) duration of heating pulse, Δt; and, (d) effect of background atmosphere (Movies S3-S10).

We have run simulations with Δt either 10 s or 30 s, since we want to reproduce variability on small temporal scales. For these short heating durations the evolution of the footpoint emission roughly tracks the timescales of the heating. In the following we focus on the simulations with 10s heating duration.

## S2.1. Sensitivity of chromosphere and TR to beam parameters

In Fig. 4 we have shown the properties of the synthetic emission (lightcurves and Si IV spectra), and, for beam simulations, the distribution of the heating as a function of temperature for a few simulations using a cool low density loop as initial condition. The simulations indicate that the TR emission is extremely sensitive to the parameters of the beam heating. For instance, the Si IV emission decreases by over an order of magnitude when increasing $E_C$ from 10 keV to 12 keV. In contrast, the Mg II k emission has a very weak dependence on both $E_C$ and $E_T$. The synthetic Mg II emission (Fig.4) rises very rapidly at the onset of the beam, because of the fast local heating of the chromospheric plasma. Some of the beam simulations show for the Mg II emission a narrow peak in the lightcurve as soon as the heating is switched off. This happens when the chromospheric layer where most of the beam energy is deposited is locally heated to temperatures larger than the characteristic formation of Mg II (Fig. S4): when the heating is switched off this plasma cools down temporarily increasing the Mg II emission. We note that our observations have 4s integration time, and therefore this late peak, which is very short lived, would not be observed with IRIS.

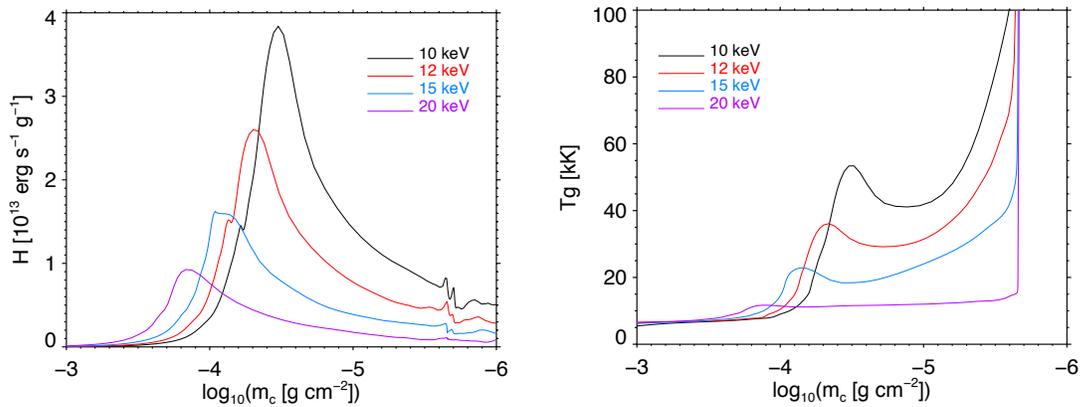

**Fig. S4**. *Heating deposition and temperature distribution in the chromosphere, for beam simulations, for simulations with different $E_C$ values.* Heating per unit mass (H, *left*) and temperature structure in the chromosphere (*right*) for RADYN loop models with different $E_C$ (10, 12, 15, 20 keV), plotted as a function of column mass density. We show the heating profiles, and the temperature profiles at the end of the heat pulse (t~10 s). For all the shown RADYN simulations the heating duration is 10 s and the $E_T = 9 \times 10^{24}$ erg.

### S2.2. Hotter and denser atmosphere

If the initial loop has higher coronal temperature and density (5 MK, $10^9$ cm$^{-3}$ respectively) the plasma emitting *AIA* 171 Å, 193 Å, and *IRIS* Si IV is confined to the small transition region layer, and at high densities (~ $10^{10}$ cm$^{-3}$).

#### a. Conduction

For conduction cases, for $E_T \lesssim 10^{25}$ erg, the increase of the temperature of the coronal plasma steepens the temperature gradient in the TR, and squeezes the volume occupied by TR plasma (see Movie S4). Even if the heating produces a density increase, the overall effect is of either a decrease of the TR emission (especially *AIA* 171 Å, 193 Å) or a very small increase in intensity, significantly lower than observed (top left panel of Fig. S5). The TR emission subsequently increases but on the longer timescales (>100 s) of the gentle chromospheric evaporation triggered by the heating event. For larger $E_T$ values ($5 \times 10^{25}$ erg) the TR emission starts increasing faster and by similar orders of magnitude of the observed brightenings (bottom left panel of Fig. S5), but the timescales are always significantly longer than observed (and than the heating duration). Also for this hotter and denser initial condition, if the TR is heated by conduction the Si IV emission is always red-shifted when the intensity its highest (panel D of Fig.4, and Fig. S5).

#### b. Beam simulations

For beam simulations, the larger coronal density implies that more of the beam energy will be deposited at larger heights than in the corresponding cases with a less dense initial atmosphere (see Movies S9 and S10). Therefore, especially for low $E_C$ values the TR will at least partly be heated by conduction from the above coronal plasma, because the soft beam will deposit much of its energy in the corona with subsequent thermal conduction transporting that energy to the TR. For $E_T \lesssim 10^{25}$ erg, and $E_C \lesssim 10$ keV, the evolution of the TR is similar to the just described conduction cases, i.e., TR emission initial decrease and subsequent evolution of long timescales (top right panel of Fig. S5). Also analogous to the conduction cases, for these beam simulations with lower $E_C$ values the Si IV emission can become redshifted (bottom right panel of Fig. S5).

For larger $E_C$ values ($\gtrsim 12$ keV), less of the beam energy is lost in the corona, and consequently the TR lightcurves are characterized by intensity increases and timescales more similar to the observations, as in the lower density case but the needed energies are larger ($E_T \gtrsim 2 \times 10^{25}$ erg).

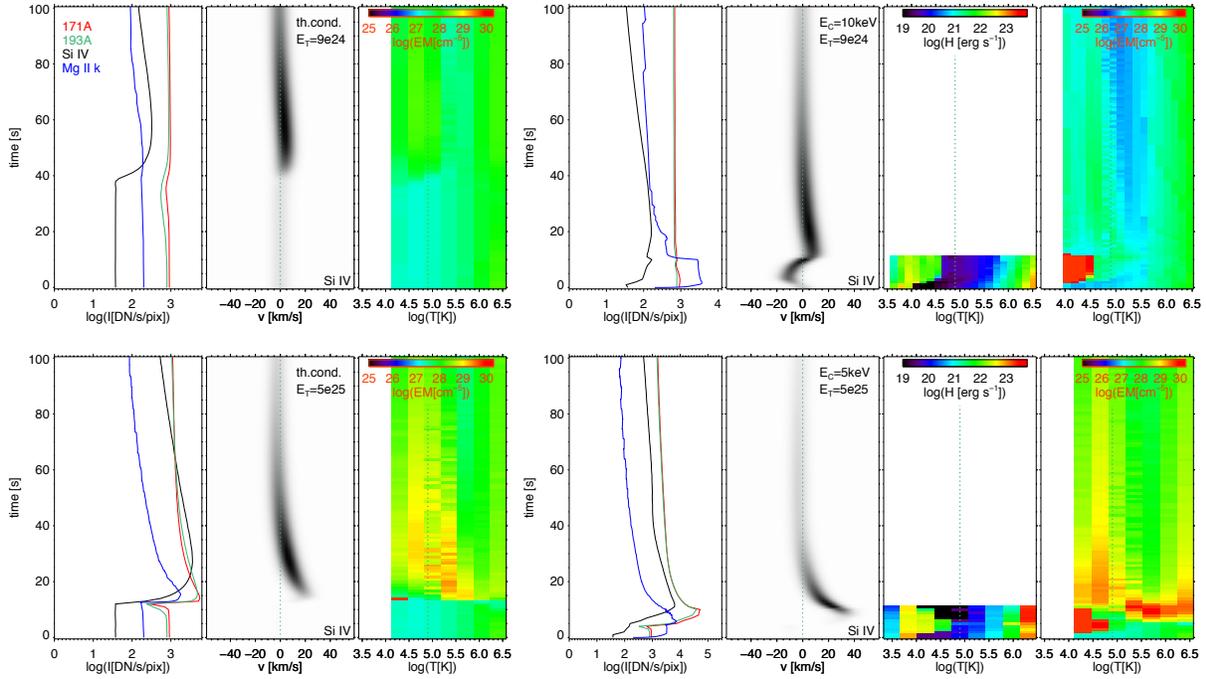

**Fig. S5.** *Modeled loop footpoint brightenings for hot and dense initial atmosphere.* Synthetic *IRIS* and *AIA* observations from RADYN simulations with heating duration of 10s. The initial atmosphere is characterized by coronal temperature of ~5 MK, and coronal density of ~$10^9$ cm$^{-3}$. The left column shows models without accelerated particles for two different $E_T$ values ($9 \times 10^{24}$ erg, top left; $5 \times 10^{25}$ erg, bottom left). The right column shows beam models with $E_C$ and $E_T$ values of: 10 keV, and $9 \times 10^{24}$ erg (top panel); and 5 keV, and $5 \times 10^{25}$ erg (bottom panel). For each model the first two panel are analogous to Fig. 4 and show respectively (1) *IRIS* and *AIA* emission lightcurves, and (2) Si IV spectra vs. time. The third panel for the beam models is also analogous to Fig. 4 and shows the heating rate as a function of temperature and time. In an additional panel (fourth panel for the beam simulations, and third panel for conduction simulations), we show the emission measure (EM(T) = $n_e^2$(T) × V(T), where $n_e$ and V are the electron density and the volume occupied by plasma at temperature T) as a function of temperature and time. The vertical dotted line in the plots of EM and H indicate the temperature of maximum formation of the Si IV 1402.77 Å line.

### S2.3. Observations of initial conditions

The *AIA* observations of the *moss* (i.e., the upper TR layer of high-pressure loops in AR) before the brightenings begin can provide some clues about the initial condition of the loops in which the footpoint brightenings occur. In Fig. S6 we show a *AIA* 171 Å image of the moss region where the footpoint brightenings are observed, but at an earlier time (~15 min before the first brightenings are observed). We superimpose the contours of the brightenings as observed by *IRIS* around 12:15UT (left panels of Fig.1 and Fig.2).

The 171 Å emission at the locations where the brightenings are observed is, for the most part, significantly weaker than the brightest patches of moss, indicating that the overlying loops are not the highest pressure loops of the active region (*11*). This, together with the results of the modeling discussed above, supports the idea that the short-lived footpoint brightenings do not occur in hot, dense loops.

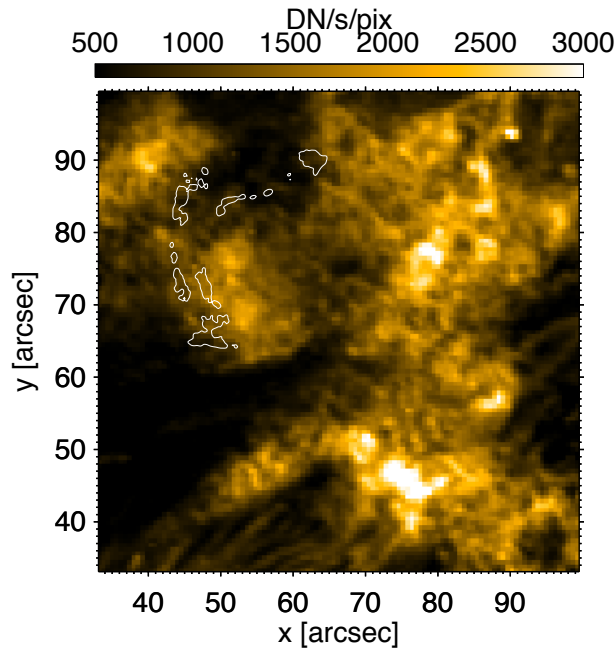

**Fig. S6**. *Moss emission before the footpoint brightenings occur.* AIA 171 Å image at 11:44:59UT, before the chromospheric/TR brightenings are observed. Contours of the *IRIS* observed brightenings (left panels of Fig.1 and Fig.2) are superimposed, and show that the brightenings are observed in moss location with much weaker emission than the brightest moss patches.

### S3. Discussion

The role of flares in coronal heating is highly debated. The similarity of flares ranging from micro- to X-class holds over many orders of magnitude, and it has been described by the "avalanche" models (*64-65*). These studies have found that the frequency of flares fits a power-law distribution with lower energy events becoming increasingly frequent. In fact, Parker (*4*) suggested that nanoflare events may be frequent enough to power the corona. The energy released by nanoflares ($\leq 10^{25}$ ergs) is below the limit of current instruments to directly detect from the flare-accelerated particles (e.g., *RHESSI*, *66*). The potential detection of hard X-rays from nanoflares by the NuSTAR mission (*67*), which is significantly more sensitive than RHESSI, has yet to be determined.

Recent high-resolution *Hi-C* observations of footpoint "moss" have shown rapid (10 – 30 s) variations in intensity that are consistent with nanoflare heating (*12*). However, the diagnostics of *Hi-C* observations were limited by the lack of spectral data, and the single band of the observations.

With *IRIS* we have observed moss brightenings on similar timescales as observed with *Hi-C*, and have analyzed the FUV and NUV high resolution spectra available for some of the brightenings. The *IRIS* data, and in particular the observed blue-shifted Si IV brightenings, indicate that at least part of the energy is deposited in the chromosphere rather than conducted down from the locally heated corona. We find that heating by non-thermal electron beams naturally reproduces the observed Si IV blueshifts as well as the range of observed intensity increases.

Chromospheric reconnection could in principle provide an alternative explanation for the observed chromospheric and TR variability, but we find that the observations support the hypothesis of beam heating. The moss brightenings clearly occur at conjugate footpoints of hot loops undergoing heating, and there is a clear correlation between the coronal and chromospheric/TR emission, naturally explained by beam heating. The Si IV brightenings are strong and occur throughout the region of the hot loop footpoints; if they were caused by chromospheric nanoflares, the reconnection and energy release would have to happen in all these locations consistently at a specific height appropriate to yield strong (and blueshifted) Si IV emission (i.e., if they occurred over a range of heights, some of them would happen too deep and would not produce any Si IV increase). Beam heating naturally explains the spatial and temporal coherence of various brightenings throughout the field-of-view, especially since the deposition height of electron beams (through the thick-target mechanism) naturally occurs at the height of the *IRIS* observations. Finally, given that moss variability is observed only at time when the overlying coronal loops are heated, if chromospheric nanoflares were the source of the observed variability, the correlation with the coronal emission would have to be explained.

By the same token, Alfven waves, proposed as an alternative energy transport mechanism, other than electron beams, from the corona to the chromosphere (*24*) would deposit most of their energy deeper in the temperature minimum region (*68*), which is again incompatible with our observations.

The *IRIS* observations strongly suggest that non-thermal particles can be generated also in very small heating events, and constrain the properties of the distributions of accelerated particles. These results can help constrain models for particle acceleration in the solar corona. The observations also provide further evidence that at least the hotter loops in active regions are likely heated impulsively (e.g.,*12,69-71*). Observing nanoflare-size events is generally challenging in coronal observations, therefore making these TR and chromospheric observations particularly valuable as a tool to investigate the characteristics of the coronal energy release and mechanism of energy transport, the response of the plasma to very impulsive heating events, and the role of particle acceleration in active region heating.  By using a combination of EUV spectral and imaging data the energy transport mechanism in flares has recently been investigated for microflares characterized by larger energy, spatial, and temporal scales (> several minutes; e.g, *72,73*) than the small events studied here.

Direct heating in the corona and subsequent energy transport to the lower atmospheric layers by thermal conduction, possibly coexists with non-thermal beam heating in the events we investigate here. The redshifts and no shifts observed for a few brightenings might be explained by either conduction or by beams with low $E_C$ value ($\lesssim$ 5 keV). We note that while the importance of non-thermal electrons in heating the *coronal* plasma in these events is not directly determined, the coronal observations are compatible with the hypothesis of the beams playing a significant role in the energetics of the coronal plasma in these loops. The evolution of the coronal emission is on significantly longer timescales than those of the initial TR and chromospheric response to the impulsive heating causing the moss variability we studied here. Therefore, the coronal emission is likely due to the contribution of many different loops heated at slightly different times. The number of impulsively heated loop aligned along the line of sight in the coronal portion of the loops is however a parameter difficult to determine from the observations, therefore hampering a quantitative forward modeling of the coronal observations.

We note that the main result of our investigation, i.e., that *IRIS* spectra of loop footpoint brightenings provide diagnostics for the presence and properties of non-thermal particles, is robust. The quantitative details of the diagnosed parameters of the non-thermal particle distributions however can have some dependence on the assumptions we have made. For instance, the exact values of $E_C$ reproducing the observations has some dependence on the spectral index $\delta$ of the electron distribution. Microflares have been found to have generally steeper power-laws ($\delta \sim 5\text{-}10$; *74*) than flares. The value of $\delta$ we assumed is therefore at the lower end of this derived range. We note, however that, if for spectral index 5 we found a very sensitive dependence on $E_C$, steeper distributions would enhance the sensitivity to $E_C$ as they would concentrate even more of the non-thermal energy around the cutoff value. Here we also adopt the assumption, generally made in analysis of hard X-ray emission from solar flares/microflares (e.g., *8, 9, 22, 74, 75*), that the electron distribution can be described by a power law with a sharp low energy cut off, while alternative distributions are possible (such as e.g., kappa distribution, *76,* or a distribution with a quasi-thermal component smoothly transitiong to a non-thermal tail, *25*).

Finally, we have calculated the expected bremsstrahlung emission from our simulations, and folded them with the *RHESSI* responses. We find that the largest expected emission corresponds to about 0.1 count in 10 s in a single detector, or about 1 count in all 9. Taking into account the background, which is of the order of 5-10 counts/s/detector in the range of interest, the expected emission is roughly 1 to 2 orders of magnitude below *RHESSI* detectability (*75*).

**Movie S1:** AIA and IRIS imaging observations of AR 11890 on 2013 November 9, 12:04:16-12:17:17, at ~12s cadence. Left to right: AIA 94 Å, AIA 193 Å passbands, IRIS slit-jaw imaging (SJI) observations in the 1400 Å passband. The AIA and IRIS observations are characterized by ~0.6"/pixel, and ~0.166"/pixel respectively.

**Movie S2:** AIA observations of AR 11890 on 2013 November 9, 11:45-12:45, at ~12s cadence, and ~0.6"/pixel. Left to right: 94 Å, 193 Å, 193 Å, 304 Å.

**Movie S3:** Results of simulations for an initial cool low density loop (T~1 MK, $n_e$~3 × $10^7$ cm$^{-3}$ in the corona), heating duration of 10 s, without non-thermal particles, and for different total energy $E_T$ (1, 3, and 6, × $10^{24}$ erg). As in Fig. S5 we show: synthetic *IRIS* and *AIA* observables (left), Si IV spectra vs. time (middle), EM as a function of temperature and time (right).

**Movie S4:** Results of simulations for an initial hot loop (T~5 MK, $n_e$~$10^9$ cm$^{-3}$ in the corona), heating duration of 10 s, without non-thermal particles, and for different total energy $E_T$ (6, 9, and 50, × $10^{24}$ erg). Plots as in Movie S3.

**Movie S5:** Results of simulations for an initial cool low density loop (T~1 MK, $n_e$~3 × $10^7$ cm$^{-3}$ in the corona), heating duration of 10 s, $E_T$ = 3 × $10^{24}$ erg, and non-thermal particles with distributions characterized by different $E_C$. As in Fig. S5, for each model we show, from left to right: (1) synthetic *IRIS* and *AIA* observables, (2) Si IV spectra vs. time, (3) heating rate as a function of temperature and time, (4) EM as a function of temperature and time.

**Movie S6:** Results of simulations analogous to those shown in Movie S5, but for $E_T$ = 6 × $10^{24}$ erg.

**Movie S7:** Results of simulations analogous to those shown in Movie S5, but for $E_T$ = 9 × $10^{24}$ erg.

**Movie S8:** Results of simulations analogous to those shown in Movie S7 ($E_T$ = 9 × $10^{24}$ erg) but for heating duration of 30 s.

**Movie S9:** Results of simulations for an initial hot loop (T~5 MK, $n_e$~$10^9$ cm$^{-3}$ in the corona), heating duration of 10 s, $E_T$ = 9 × $10^{24}$ erg, and non-thermal particles with distributions characterized by different $E_C$. Plots are analogous to Movies S5-S8.

**Movie S10:** Results of simulations analogous to those shown in Movie S9, but for $E_T$ = 5 × $10^{25}$ erg.

**References:**

31. P. Boerner et al. *Solar Physics* **275**, 41 (2012)

32. J. Martinez-Sykora et al. *Astrophys. J.* **743**, 23 (2011)

33. P. Testa, E. Landi, J. Drake, *Astrophys. J.* **745**, 111 (2012)

34. A. Foster, P. Testa, *Astrophys. J. Letters* **740**, 52 (2011)

35. P. Testa, F. Reale, *Astrophys. J. Letters* **750**, 10 (2012)

36. L. Teriaca, H. Warren, W. Curdt, *Astrophys. J. Letters* **754**, 40 (2012)

37. J. Leenaarts et al., *Astrophys. J.* **772**, 90 (2013)

38. H. Carmichael, *NASA Special Publication* **50**, 451 (1964)

39. P. Sturrock, *Nature* **211**, 695 (1966)

40. T. Hirayama, *Solar Physics* **34**, 323 (1974)

41. R. Kopp, & G. Pneuman, *Solar Physics* **50**, 85 (1976)

42. L. Fletcher et al., *Space Sci. Rev.* **159**, 19 (2011)

43. A. Shih, R. Lin, D. Smith, *Astrophys. J. Letters* **698**, 152 (2009)

44. G. Emslie et al., *Journal of Geophysical Research* **109**, A10104 (2004)

45. M. Carlsson, *Astrophys. J. Letters* **397**, 59 (1992)

46. M. Carlsson, R. Stein, in *Proc. Mini-Workshop on Chromospheric Dynamics,* ed. M. Carlsson (Oslo: Inst. Theor. Astrophys.), 47 (1994)

47. M. Carlsson, R. Stein, *Astrophys. J. Letters* **440**, 29 (1995)

48. E. Dorfi, L. Drury, *J. Comput. Phys.* **69**, 17 (1987)

49. K. Dere et al., *Astron. & Astrophys. Suppl.* **125**, 149 (1997)

50. E. Landi et al., *Astrophys. J.* **763**, 86 (2013)

51. D. Smith, L. Auer, *Astrophys. J.* **238**, 1126 (1980)

52. W. Abbett, S. Hawley, *Astrophys. J.* **521**, 906 (1999)

53. J. Allred et al., *Astrophys. J.* **630**, 573 (2005)

54. J. Allred, A. Kowalski, M. Carlsson, *Astrophys. J.* in preparation (2014)

55. W. Liu, V. Petrosian, J. Mariska, *Astrophys. J.* **702**, 1553 (2009)

56. J. McTiernan, V. Petrosian, *Astrophys. J.* **359**, 524 (1990)

57. G. Peres et al., *Astrophys. J.* **252**, 791 (1982)

58. R. Betta et al., *Astron. & Astrophys. Suppl.* **122**, 585 (1997)

59. H. Uitenbroek, *Astrophys. J.* **557**, 389 (2001)

60. J. Leenaarts, T. Pereira, H. Uitenbroek, *Astron. & Astrophys.* **543**, 109 (2012)

61. M. Asplund et al., *Ann. Rev. of Astron. & Astrophys.* **47**, 481 (2009)

62. P. Testa et al., *Astrophys. J.* **758**, 54 (2012)

63. U. Feldman, *Phys. Scr.* **46**, 202 (1992)

64. E. Lu, R. Hamilton, *Astrophys. J. Letters* **380**, 89 (1991)

65. E. Lu et al., *Astrophys. J.* **412**, 841 (1993)

66. R. Lin et al., *Solar Physics* **210**, 3 (2002)

67. F. Harrison et al., *Astrophys. J.* **770**, 103 (2013)

68. A. Russelll, & L. Fletcher, *Astrophys. J.* **765**, 81 (2013)

69. F Reale et al., *Astrophys. J.* **698**, 756 (2009)

70. F. Reale et al., *Astrophys. J. Letters* **736**, 16 (2011)

71. N. Viall, & J. Klimchuk, *Astrophys. J.* **753**, 35 (2012)

72. J. Brosius, & G. Holman, *Astron. & Astrophys.* **540**, 24 (2012)

73. J. Brosius, *Astrophys. J.* **754**, 54 (2012)

74. I. Hannah, et al., *Astrophys. J.* **677**, 704 (2008)

75. I. Hannah, et al., *Astrophys. J.* **724**, 487 (2010)

76. M. Oka, et al., *Astrophys. J.* **764**, 6 (2013)